%% file: main.tex
\documentclass{article}

\usepackage{url}
\usepackage{cite}
\usepackage{soul}
\usepackage{ulem}
\usepackage{arxiv}
\usepackage{xspace}
\usepackage{siunitx}
\usepackage{graphicx}
\usepackage{textcomp}
\usepackage{graphicx}
\usepackage{multirow}
\usepackage{booktabs}
\usepackage{placeins}
\usepackage{subcaption}
\usepackage{algorithmic}
\usepackage[table,xcdraw]{xcolor}
\usepackage{amsmath,amssymb,amsfonts}

\def\BibTeX{{\rm B\kern-.05em{\sc i\kern-.025em b}\kern-.08em
    T\kern-.1667em\lower.7ex\hbox{E}\kern-.125emX}}
\begin{document}
\title{From Vision to Sound: Advancing Audio Anomaly Detection with Vision-Based Algorithms}

\author{
    Manuel Barusco \\ 
    University of Padova, Italy \\ 
    \texttt{manuel.barusco@phd.unipd.it} \\ \And 
    Francesco Borsatti \\ 
    University of Padova, Italy \\ 
    \texttt{francesco.borsatti.1@phd.unipd.it} \\ \And
    Davide Dalle Pezze \\ University of Padova, Italy \\ 
    \texttt{davide.dallepezze@unipd.it} \\ \And
    Francesco Paissan \\ Italy \\ 
    \texttt{francescopaissan@gmail.com} \\ \And
    Elisabetta Farella \\ Fondazione Bruno Kessler, Italy \\ 
    \texttt{efarella@fbk.eu} \\ \And
    Gian Antonio Susto \\ 
    University of Padova, Italy \\ 
    \texttt{gianantonio.susto@unipd.it} \\
}

\maketitle

\begin{abstract}
Recent advances in Visual Anomaly Detection (VAD) have introduced sophisticated algorithms leveraging embeddings generated by pre-trained feature extractors.
Inspired by these developments, we investigate the adaptation of such algorithms to the audio domain to address the problem of Audio Anomaly Detection (AAD).
Unlike most existing AAD methods, which primarily classify anomalous samples, our approach introduces fine-grained temporal-frequency localization of anomalies within the spectrogram, significantly improving explainability. This capability enables a more precise understanding of where and when anomalies occur, making the results more actionable for end users. 
We evaluate our approach on industrial and environmental benchmarks, demonstrating the effectiveness of VAD techniques in detecting anomalies in audio signals.
Moreover, they improve explainability by enabling localized anomaly identification, making audio anomaly detection systems more interpretable and practical.
\end{abstract}


\section{Introduction}

Audio anomaly detection (AAD) is the task of detecting unexpected or out-of-distribution sounds in audio sequences. This task finds many real-world applications, especially in the industrial domain, e.g., for quality control and predictive maintenance \cite{deschenes2025planing,folz2024investigation,kilickaya2024audio}. 
Beyond industry, AAD also plays a key role in environmental surveillance, enhancing public safety through the detection of anomalous acoustic events in urban areas \cite{bajovic2021marvel, melgar2023identification, jalillarge}. 
Recently, AAD systems using deep neural networks have achieved remarkable accuracy \cite{abbasi2022large,deschenes2025planing,purohit1909mimii}. In \cite{koizumi2020description}, the authors proposed using an autoencoder.
While more recent methods are proposed in literature \cite{kuroyanagi2025serial,deschenes2025planing}, usually based on the autoencoder concept, a major limitation of existing methods is their lack of explainability. These models operate as black boxes, providing anomaly scores without explaining which parts of the audio are responsible for the anomaly detection. This opacity limits their usability in critical applications where understanding the decision-making process is essential.
Meanwhile, significant advancements have been made in Visual Anomaly Detection (VAD). In particular, most recent algorithms propose to leverage embeddings from pre-trained feature extractors to detect anomalies on each patch in the feature map \cite{patch}, as explained in Section \ref{subsec:considered_vad_approaches}. Therefore, they can predict anomaly scores at the pixel level; this makes them more explainable, as they highlight the portions of the input images deemed out-of-distribution, commonly referred to as \emph{anomaly maps}. These maps resemble the saliency maps used in recent audio explainability literature \cite{paissan2024listenable}. Given their high explainability and practical value, such techniques are highly valuable in real-world applications where understanding why an input is anomalous is as important as detecting the anomaly itself.

\begin{figure}[!bth]
    \centering
    \includegraphics[width=0.7\linewidth]{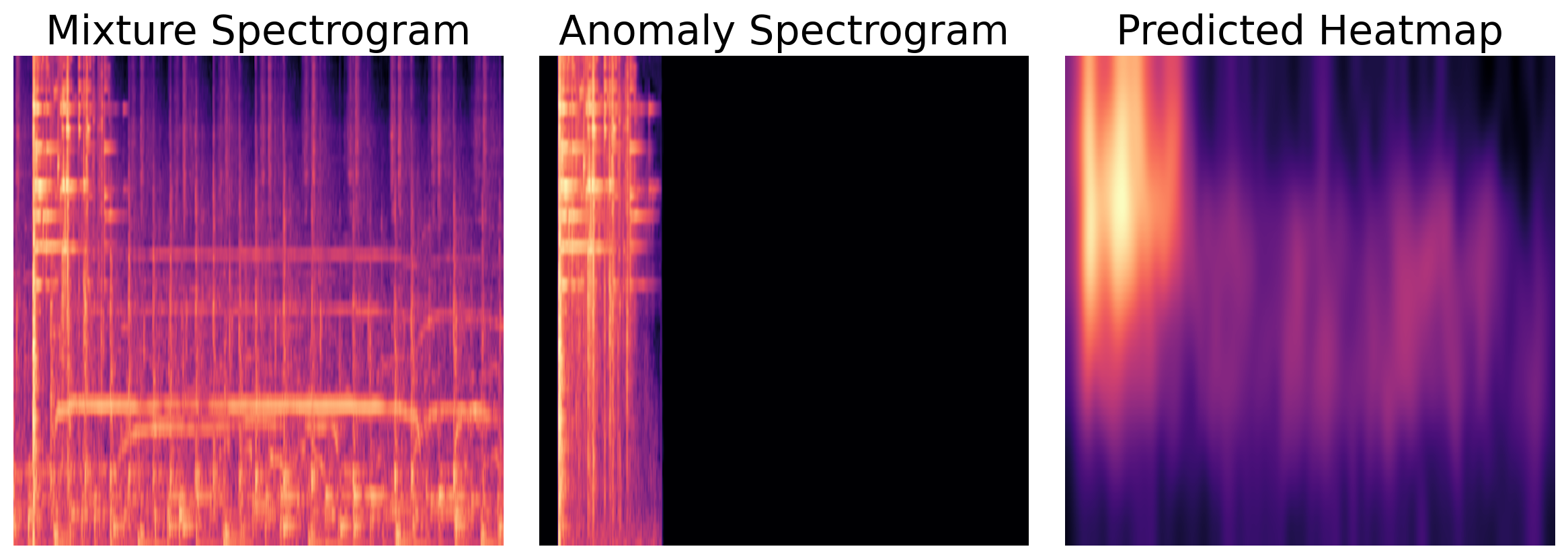}
    \caption{\textbf{Example of interpretability results obtained by our Approaches} The first image illustrates a mixed (corrupted) audio signal containing both normal and anomalous components. The second image displays the spectrogram of the anomaly alone. The third image shows the anomaly map generated by the model, highlighting its ability to accurately identify and isolate the anomalous segments within the mixed signal.
    }
    \label{fig:interpretability_example}
\end{figure}

Building on these insights, we explore the adaptation of VAD techniques to the audio domain, investigating their effectiveness in AAD and their ability to generate meaningful and explainable anomaly maps. Unlike most existing AAD methods, our approaches not only detect anomalous samples but also provide fine-grained temporal-frequency localization of anomalies within the spectrogram (Fig. \ref{fig:interpretability_example}). This feature significantly improves explainability, providing users with clearer insights and facilitating more informed decision-making.

As an additional contribution, we also propose a novel set of evaluation metrics to assess how effectively algorithms identify abnormal regions in spectrograms, addressing a significant gap in the current literature where such evaluation methods are (to the best of our knowledge) absent.

\section{Methodology}
\label{sec:Methodology}

\subsection{Framework}
\label{subsec:Framework}

\begin{figure*}[thbp]
  \centering
    \includegraphics[width=0.8\textwidth]{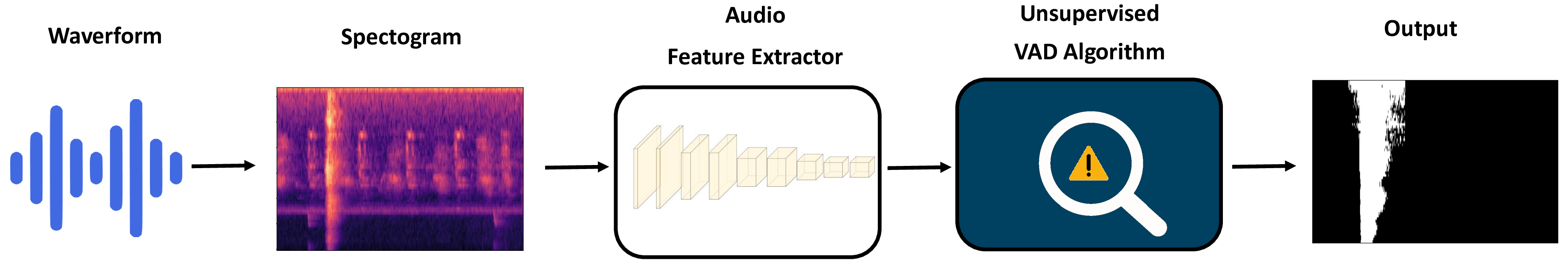}
    \caption{ We propose to work on the embeddings produced by the audio feature extractor and test several sota algorithms proposed originally for Visual Anomaly Detection
    }
    \label{fig:framework_AAD_embeddings}
\end{figure*}

The following section describes the framework adopted to evaluate the VAD algorithms in the audio domain performing anomaly detection in the unsupervised setting. The framework is presented in Fig. \ref{fig:framework_AAD_embeddings}.
The practical advantage of considering an unsupervised setting is eliminating the need for an expensive and time-consuming process to acquire and label (rare) abnormal samples.

Most of the current research in the field of VAD is focused on \textbf{feature-based methods}.
Despite each algorithm processing the input images differently, most of them work on the embeddings produced by a pre-trained feature extractor. The advantage of using a pre-trained model is that the obtained embeddings are generic and capture high-level features that are usable across several contexts. 
Moreover, leveraging pre-trained models has the practical advantage of reducing computational costs, as they eliminate the need for expensive model training.
Furthermore, when training a model from scratch using domain-specific audio data, the model may struggle to generate sufficiently rich representations that can be used effectively.

A peculiarity of VAD methods is that a generic feature map $H \times W \times C$ can be split into $H \cdot W$ vectors of dimension $C$, and usually, their abnormality is assessed separately for each vector.
Each of these vectors is associated with a specific region (or patch) of the input image.
In other words, when operating at the patch level, these methods focus on small regions of the image rather than the image as a whole, which helps the model identify abnormal regions more easily and, by extension, classify the entire sample.

Moreover, this helps to improve the explainability since the model can not only determine whether a sample is abnormal but can also identify the specific regions of the image that are considered suspicious (see Fig \ref{fig:interpretability_example} for an example).

Given the many advantages of VAD algorithms, their adoption in the audio domain can have a significant impact.
Therefore, we focus on applying features-based VAD methods to the audio domain.
In our framework, after transforming the waveform into an adequate spectrogram, the produced spectrogram is used as input to a feature extractor to obtain an intermediate generic representation.
In our process, an appropriate pre-trained model for the audio domain is selected to obtain good embedding representations (see Sec. \ref{subsec:models}).
These embeddings are then used to train the VAD algorithms (a brief description of each tested method is provided in Sec. \ref{subsec:considered_vad_approaches}), and the resulting output is used as a heatmap of the spectrogram, which indicates whether the sample is abnormal and highlights the regions contributing to the model's prediction.
The entire process is illustrated in Fig. \ref{fig:framework_AAD_embeddings}.

\input{tables/table_results}

\input{tables/table_ff}

\subsection{Considered VAD Approaches}
\label{subsec:considered_vad_approaches}
Usually, the feature-based VAD algorithms are commonly categorized into four distinct groups: \textbf{memory-bank approaches, teacher-student approaches, one-class classification methods, and normalizing-flow approaches}. 
For each tested method, we provide a brief description below:
\\
\textbf{PatchCore}: This is a memory-bank method that uses additional memory to store a set of patches embeddings \cite{patch}. 
During inference, a nearest-neighbor lookup is applied to calculate the anomaly score of each input image patch.
\\
\textbf{Padim}: This is also a memory-bank method. 
However, instead of storing the patches embeddings, Padim proposes that each patch position in the input image can be characterized by a multivariate Gaussian distribution.
During the inference process, for a given test image, the Mahalanobis distance is computed for each patch, providing its anomaly score \cite{PaDiM}.
\\
\textbf{CFA}: This is a memory-bank approach that, given the patch feature vectors, trains a Patch Descriptor Network (PDN) to increase the density of normal features vectors and make it easier to separate normal and abnormal patches \cite{lee2022cfa}.
\\
\textbf{STFPM}: This approach belongs to the student-teacher category.
As the name suggests, it utilizes an architecture comprising two separate models: a student (randomly initialized) and a teacher (pre-trained).
During training, the student aims to produce the same embeddings as the teacher for the normal samples patches, causing during inference high differences when examining abnormal samples \cite{st_pyramid}.

\input{tables/table_results_env}

\section{Experimental Setting}
\label{sec:experimental_setting}

\subsection{Benchmarks}
\label{subsec:datasets}

In the current research literature, AAD is considered for two types of applications.

The first is industrial applications, where AAD can be used to detect equipment failures or abnormal behaviors. 
The second is environmental applications, where AAD can enhance urban safety or aid in environmental disaster prevention. 

While previous studies in the literature tend to examine these two applications independently, our approach evaluates all the VAD models presented in Section \ref{subsec:considered_vad_approaches} across both domains.
This allows us to offer a more comprehensive assessment of VAD algorithms within the audio domain, better understanding the generalizability of these models.

\textbf{MIMMI Benchmark}: To evaluate the performance in the industrial domain, we consider the most used dataset in literature called MIMMI (Malfunctioning Industrial Machine Investigation and Inspection) Dataset \cite{purohit1909mimii}.
It includes recordings from four types of industrial machines: \textit{valves, pumps, fans, and slide rails}.
Each machine type features four individual product models, with data comprising both normal and anomalous operating sounds. The anomalies simulate real-world issues such as contamination, leakage, rotating imbalance, and rail damage. 
Moreover, to closely represent factories, the normal audio is mixed with background noise from multiple real factories. Our experiments focus on all the categories, machine IDs, and noise levels present in the dataset.

\textbf{EnvMix Benchmark}: We also study the case where we want to perform AAD on environmental audio that corresponds to different domains such as city, room, etc.
We draw inspiration from current literature that mixes normal signals with anomalous signals  \cite{conte2012ensemble,jalillarge,mesaros2017dcase}.
Specifically, the creation of the dataset is obtained by mixing environmental signals (normal audio) with anomalous sounds that represent unexpected behavior.
Specifically, we consider as normal signals the following list of categories from the UrbanSound8K dataset \cite{salamon2014dataset}: \textit{air conditioner, engine idling, street music, drilling, car horn, jackhammer, children playing, siren}.
We considered the ESC50 dataset \cite{piczak2015dataset} to select a series of categories to represent the anomalous sounds injected in the normal audio: \textit{glass breaking, coughing, sneezing, door wood knock, rooster, can opening, car horn, dog, laughing, door wood creaks, clock alarm, clapping, sheep, snoring, fireworks, washing machine, frog}.

Our evaluation will be performed for both benchmarks considering different levels of SNR (factor noise): 6 dB, 0 dB, and -6 dB.
This means that we are going to measure the ability of the models by varying the difficulty level of the anomaly detection problem, where higher noise levels make anomaly detection more challenging.

For additional results and additional details, you can refer to the repository \footnote{ 
\url{https://bitbucket.org/papers_vad_group/vision_to_sound_ad/src/main/}}.

\subsection{Metrics}
\label{subsec:metrics}

Various evaluation metrics are commonly employed to assess the performance of AD techniques.

\subsubsection{Sample-level}
Traditional evaluation of AD techniques relies on multiple metrics to provide a comprehensive assessment of performance.
At the sample level, two metrics are the most commonly employed: F1 and ROC.
While ROC is useful since it is independent by a specific threshold, it is much less robust in the presence of imbalanced datasets, which is common in anomaly detection where the normal class is more frequent than the anomalous one.

\subsubsection{Temporal-Frequency Localization}

One of the key contributions of this work is to evaluate VAD algorithms for AAD beyond the sample-level analysis and focus on spectrogram analysis, enabling the detection of the specific anomalous regions of the spectrogram.

We introduce novel metrics to assess how effectively algorithms identify abnormal regions in spectrograms, addressing a significant gap in the current literature where such evaluation methods are (to the best of our knowledge) absent.

Given the anomaly map produced by a VAD approach (see Fig. 4.b for an example) and the ground truth, the F1, ROC and AU-PRO \cite{Bugarin_2024_CVPR} metrics are calculated.
The ground truth mask of the anomalous spectrogram is created by identifying the top 40\% most energy-intensive values in the spectrogram area where the anomalous sound is injected.

We define an anomaly map $M \in [0,1]^{H \times W}$,
where $M_{t,f}$ indicates the probability, according to the model, that the temporal instant $t$ is anomalous in the set of frequencies $f$.
Anomalies are identified as values $M_{f,t}$  exceeding the 40th percentile $p_{40}$
of the spectrogram's value distribution. Formally, the predictions are defined as:
$G(t,f) = \mathbb{I}_{\{M_{t,f} > p_{40}\}}$, where $\mathbb{I}$ is the indicator function.

\subsubsection{Temporal Localization}
While the previous metrics allow the end user to achieve a high comprehension level of where the anomalous regions are located inside the spectrogram, they could be excessively informative.
Therefore, we also evaluate the ability of the model to provide where the anomaly is present only on the temporal axis. 
Specifically, the model identifies all temporal instants where an anomaly occurs.
To define the ground truth on the temporal axis, for each temporal instant where the anomaly sound is injected, we sum the values of all the frequencies
$ E_t = \sum_{f=1}^{F} \log(M_{t,f})$.
Then, we define as anomalous all the temporal instants where the energy is above the 50-th percentile, formally: 
$\text{Anomaly}(t) = \mathbb{I}_{\{E_t > p_{50}\}}$
where $p_{50}$ represents the 50th percentile of the values of the anomalous spectrogram and $\mathbb{I}$ is the indicator function.
Given the model anomaly map $M$, we identify the anomaly score of a temporal instant $t$ as the average of the five most anomalous values in $M_{t}$.
Using the ground truth and predictions as defined, we calculate the ROC and F1 metrics.

\subsubsection{Faithfulness}
In addition, we evaluate the ability of the model to isolate the audio signal from the input.
This helps to understand how well the model identifies the anomalous part of the spectrogram.
To achieve this, we calculate the filtered audio as follows: 
$\tilde{x} = x \odot M$.
Next, we assess the quality of the obtained signal using the metric \textbf{Faithfulness on Spectra (FF v1)} as defined in \cite{parekh2022listen}.
The metric is calculated by measuring the difference in model probability predictions (at the sample level) before and after the filtering.
More formally, let's define the model as $f$ where $f(x)$ produces the sample-level prediction of a generic sample $x$.
Then we calculate FF as follows: $FF = f(x) - f(\tilde{x})$.
If the value is high, the prediction changes significantly, transitioning from a highly anomalous signal to a normal signal (or vice-versa).
Therefore, if FF is high, the filtered signal was masked correctly, effectively removing the anomalous part from the contaminated signal.
Since $\tilde{x}$ can have holes in the spectrogram caused by areas in $M$ with very low anomaly scores, we formalized the \textbf{Faithfulness on Spectra v2 (FF v2)} which is calculated as before but considering $\tilde{x} = x \odot (1 - M) + bg \odot M$
where $bg$ is the background sound used for generating the contaminated signal. $bg$ is used for covering the "holes" in the spectrogram.

\subsection{Models}
\label{subsec:models}

As discussed in Sec. \ref{sec:Methodology}, our tested methods 
perform on embedding produced by a pre-trained feature extractor.
This means that they can exploit generic and high-level representations produced by models pre-trained on large datasets of similar domains. 
However, this also indicates that the quality of the representations produced by the chosen feature extractor significantly influences the performance of the VAD algorithms.
In other words, selecting the most correct feature extractor is an extremely relevant phase in obtaining good results.

In our work, we considered as a feature extractor the CNN14 from the CLAP model \cite{elizalde2023clap}.
CNN14 is a convolutional neural network trained in a contrastive learning framework, where the model learns to match audio content with corresponding text descriptions.

From previous studies in VAD, the optimal choice of layers is a set that includes low-level, medium-level, and high-level layers \cite{barusco2024paste}.
This is because different layers represent different levels of resolution, and a representation that combines different resolutions helps to look at both general and specific features.
Therefore, we similarly select a set of layers that provide information on different resolution levels.
Specifically, we select the layers $conv\_block2$, $conv\_block3$, $conv\_block4$
We compare the VAD models with a \textit{Baseline} introduced in \cite{koizumi2020description}, which is a simple autoencoder.

\section{Results}
\label{section:Results}

\subsection{Performance}
\label{subsec:performance_results}

We begin by examining the results for the industrial domain (MIMII Dataset) and then continue examining the results for the environmental domain (EnvMix Dataset).

In the MIMMI Dataset, we compare the baseline provided in \cite{dcase} with our four tested VAD algorithms: CFA, Padim, PatchCore, and STFPM.
Our results show that in three out of the four categories, the VAD algorithms show superior performance than the baseline.
This is encouraging, demonstrating the potential of adapting VAD methods to the audio domain and suggesting that future work on these methods could help to increase the performance further.
In particular, the best method seems to be STFPM, with the best value for two out of the four categories.

When examining the performance of VAD algorithms for EnvMix Dataset, we can observe that the best method among the four is PatchCore.
On one hand, this is not surprising since PatchCore is the state-of-the-art for Computer Vision.
On the other hand, it is interesting that PatchCore wasn't the best-performing method for any categories of the MIMII Dataset.
This suggests that the underlying domain could influence the decision on the best method to use.

In particular, it is interesting that PatchCore shows good results not only at the sample level with ROC 90.6 but also at the spectrogram level with ROC 76.4 and at the temporal level with ROC 82.9.

\subsection{Explainability Results}
\label{subsec:overall_analysis}

As stated above, PatchCore performs well on audio signals of the EnvMix Dataset with high values of ROC for sample, spectrogram, and temporal levels.
In other words, PatchCore obtains good results in identifying not only the presence of anomalies in the audio signal but also the localization of the anomaly in the spectrogram and which temporal instants are associated with the anomalous signal.
We can examine this deeply by visual examination, as shown in Fig. \ref{fig:interpretability_example}.
Here, the mixed signal is shown in the first plot, while in the second plot, the anomalous signal is added to the normal signal to obtain the mixed audio.
Then, the third figure shows the anomaly map produced by one of the VAD algorithms.
We can see visually how the model detects the part where the anomaly is present well.

Regarding the \textit{Faithfulness} metric, from the results reported in Tab. \ref{tab:ff} we can see that the obtained values are quite low. This is caused by the fact that the chosen AD models are very sensitive to every change in the input spectrogram: every deviation from the normal input distribution is anomalous. We tried two versions of the metric that change the input spectrograms in two different ways. The FF v2 tries to tackle the problem by producing less anomalous spectrograms.
However, in both cases, when the model produces a not-accurate anomaly map, the obtained spectrogram can have some artifacts that can trigger the model again.

\section{Conclusion}
\label{sec:conclusion}

The results demonstrate the capacity of the VAD methods to identify anomalous samples, 
showing the potential of adapting VAD techniques to the audio domain.

Moreover, this is achieved while enabling more explainable and practical audio anomaly detection systems by localizing the anomalous regions of the spectrogram.
Unlike most existing approaches in the AAD literature, our approaches detect anomalous samples and provide fine-grained temporal-frequency localization of anomalies within the spectrogram.
This added level of detail significantly enhances the explainability of the results, making them particularly valuable for informed decision-making by end users.

While the obtained results show the potential of VAD algorithms in the audio domain, future work is required to improve the methods' performance to accompany the explainability results with strong detection performance.

A promising future research direction is to explore the role of feature extractors by testing different models.
This is because, for the tested VAD algorithms, the performance heavily depends on the quality of the used feature representations.
In particular, it would be interesting to analyze if employing self-supervised learning methods could help the model generate more effective feature representations for VAD algorithms.

In addition, while several novel metrics to evaluate the interpretability of the methods were introduced in this work, a better formalization of the Faithfulness metric for the AAD problem remains necessary.

\bibliographystyle{IEEEtran}
\bibliography{references}

\end{document}

%% file: tables/table_results.tex
\begin{table}[htb]
\caption{MIMII benchmark results and comparison with DCASE 2020 Task 2 baseline.}
\label{tab:mimii_bench}
\centering
\begin{tabular}{lcccccccc}
\toprule
& \multicolumn{2}{c}{Fan}          & \multicolumn{2}{c}{Pump}         & \multicolumn{2}{c}{Slider}                & \multicolumn{2}{c}{Valve} \\
\textbf{Method} & F1   & \multicolumn{1}{c|}{ROC}  & F1   & \multicolumn{1}{c|}{ROC}  & F1   & \multicolumn{1}{c|}{ROC}           & F1               & ROC    \\ \midrule
Baseline        & -    & \multicolumn{1}{c|}{65.8} & -    & \multicolumn{1}{c|}{72.9} & -    & \multicolumn{1}{c|}{\textbf{84.8}} & -                & 66.3   \\
CFA &
  70.7 &
  \multicolumn{1}{c|}{65.7} &
  77.9 &
  \multicolumn{1}{c|}{76.0} &
  \textbf{81.2} &
  \multicolumn{1}{c|}{79.9} &
  75.9 &
  \textbf{73.2} \\
PaDiM           & 69.2 & \multicolumn{1}{c|}{61.1} & 72.3 & \multicolumn{1}{c|}{65.3} & 79.0 & \multicolumn{1}{c|}{78.5}          & \textbf{76.4}    & 71.8   \\
PatchCore       & 67.5 & \multicolumn{1}{c|}{59.6} & 68.7 & \multicolumn{1}{c|}{59.3} & 75.2 & \multicolumn{1}{c|}{76.2}          & 70.8             & 67.1   \\
STFPM &
  \textbf{74.0} &
  \multicolumn{1}{c|}{\textbf{72.5}} &
  \textbf{83.8} &
  \multicolumn{1}{c|}{\textbf{85.9}} &
  76.1 &
  \multicolumn{1}{c|}{74.4} &
  72.4 &
  66.3 \\ \bottomrule
\end{tabular}%
\end{table}

%% file: tables/table_ff.tex
\begin{table}[htb]
\centering
\caption{Faithfulness results for different AD methods in the EnvMix Dataset.}
\label{tab:ff}
\begin{tabular}{l|c c c c}
\toprule
 & CFA & PaDiM & PatchCore & STFPM \\ 
\midrule
FF v1 & \( 0.023 \pm 0.021 \) & \( -0.031 \pm 0.101 \) & \( -0.134 \pm 0.115 \) & \( 0.021 \pm 0.026 \) \\  
FF v2 & \( 0.035 \pm 0.022 \) & \( 0.038 \pm 0.137 \) & \( 0.062 \pm 0.151 \) & \( 0.029 \pm 0.019 \) \\  
\bottomrule
\end{tabular}
\end{table}

%% file: tables/table_results_env.tex
\begin{table}[htb]
\caption{Results for EnvDataset  for different levels of analysis (sample-level, spectrogram-level, temporal-level) }
\label{tab:anomaly_detection_comparison}
\centering
\begin{tabular}{lccccccc}
\toprule
& \multicolumn{2}{c}{Sample Level} & \multicolumn{3}{c}{Spectrogram Level}   & \multicolumn{2}{c}{Temporal Level} \\
\multicolumn{1}{l|}{Method} & ROC  & \multicolumn{1}{c|}{F1}   & F1   & PRO  & \multicolumn{1}{c|}{ROC}  & F1               & ROC             \\ \midrule
\multicolumn{1}{l|}{CFA}    & 79.3 & \multicolumn{1}{c|}{76.4} & 27.6 & 56.0 & \multicolumn{1}{c|}{64.6} & 46.5             & 70.0            \\
\multicolumn{1}{l|}{PaDiM}  & 86.3 & \multicolumn{1}{c|}{81.8} & 35.0 & 56.9 & \multicolumn{1}{c|}{72.7} & 55.0             & 78.9            \\
\multicolumn{1}{l|}{PatchCore} &
  \textbf{90.6} &
  \multicolumn{1}{c|}{\textbf{85.8}} &
  \textbf{36.8} &
  \textbf{58.3} &
  \multicolumn{1}{c|}{\textbf{76.4}} &
  \textbf{59.9} &
  \textbf{82.9} \\
\multicolumn{1}{l|}{STFPM}  & 53.2 & \multicolumn{1}{c|}{67.2} & 19.4 & 47.1 & \multicolumn{1}{c|}{47.0} & 35.4             & 47.9  \\
\bottomrule
\end{tabular}%
\end{table}